\documentclass[aps,twocolumn,floats,prx,longbibliography,nofootinbib]{revtex4-2}
\usepackage{graphics,graphicx,epsfig}
\usepackage{amssymb,color}
\usepackage{amsmath,textgreek,bm}
\usepackage{epsf,epstopdf,wrapfig}
\usepackage{bigints}

\begin{document}

\title{When many noisy genes optimize information flow}

\author{Nicholas Lawson and William Bialek}
\affiliation{Joseph Henry Laboratories of Physics and Lewis--Sigler Institute for Integrative Genomics\\
    Princeton University, Princeton, NJ 08544 USA}

\date{\today}

\begin{abstract}
It often is emphasized that gene expression is noisy.  A seemingly contradictory view is that control mechanisms have been optimized to squeeze as much information as possible out of a limited number of molecules.  Here we revisit these issues in a simple model where a single transcription factor (TF) controls a large number of target genes.  We include only the physically required noise sources---random arrival of TFs at their targets and counting noise in the synthesis and degradation of mRNA.  If the cell has a limited total number of mRNA molecules, then the capacity to transmit information about TF concentration is maximized when these resources are distributed across the largest possible number of target genes.  To realize this capacity the distribution of TF concentrations must be biased toward smaller values.  Thus, in some limits, information transmission is optimized when individual expression levels are noisy.  In addition, the dependence of information transmission on the parameters of this multi--gene system has a ``sloppy'' spectrum, so that optimal performance can co--exist with substantial variability.
\end{abstract}

\maketitle

\section{Introduction}

The macroscopic behaviors of organisms can be reliable, reproducible, and robust.  In some cases, as with the ability of our visual system to count single photons on a dark night, performance is so precise that it approaches the limits of what is allowed by the laws of physics \cite{hecht+al_42,rieke+baylor_1998}.  But as we look at the microscopic structures that underlie these macroscopic behaviors things can seem much messier:  the array of receptors in our retina is highly disordered \cite{roorda+al_01}, numbers of ion channels vary wildly from neuron to neuron even in organisms where these cells can be identified and numbered \cite{marder+goaillard_06}, and the expression levels of individual genes fluctuate substantially across genetically identical cells \cite{raser+oshea_2005}.  Although more attention is given to noisiness and disorder, there are enough examples of precision and near--optimality that we should worry about how to reconcile these different classes of observations \cite{Bialek_2012,Bialek_2024}.

It is an old idea that part of the problem in understanding the physics of life is to understand how reliable functions emerge in networks with unreliable components \cite{vonNeumann_1956}.  But, for example, neurons in the brain can give reliable responses to injected currents \cite{Mainen+Sejnowski_1995} and to less proximal sensory inputs \cite{spikesbook,Hires+al_2015}.  Similarly, the genetic networks operating in the first few hours of embryonic development in the fruit fly provide evidence for optimization in the encoding and decoding of information through the regulation of gene expression \cite{gregor+al_2007b,dubuis+al_2013b,petkova+al_2019,mcgough+al_2024}.

\begin{figure}[b]
\includegraphics[width=\columnwidth]{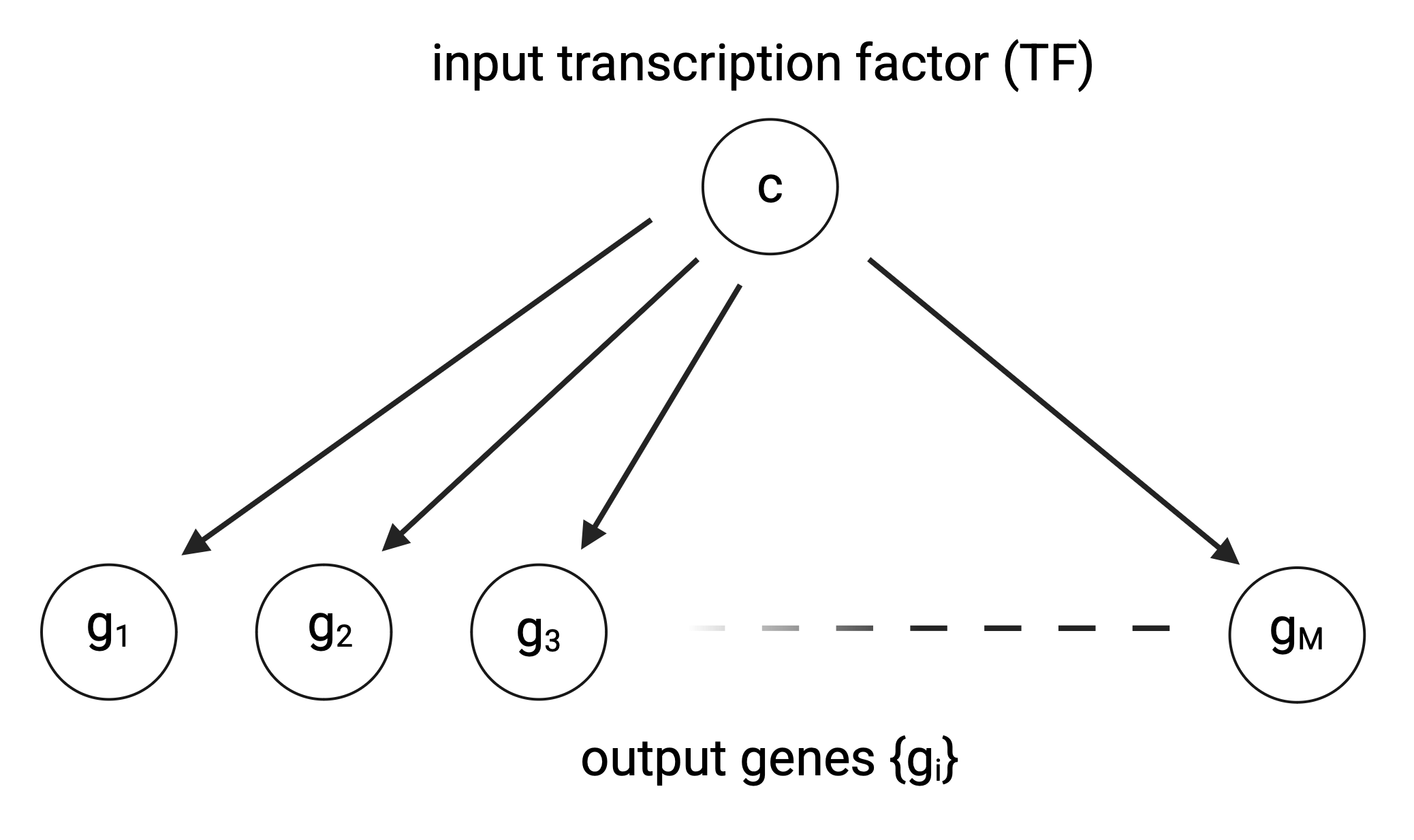}
    \caption{A schematic of the broadcasting model. One input species, a transcription factor at concentration $c$, independently regulates the expression levels $\{g_i\}$ of $M$ output genes ($i = 1,\, 2,\, \cdots ,\, M$).
    \label{fig:broadcast}}
\end{figure}

Here we develop a different scenario connecting noise and optimization in the regulation of gene expression.  We imagine a single transcription factor that can regulate many genes; for simplicity we imagine that these genes are non--interacting, so that the single input ``broadcasts'' to many outputs,  as schematized in Fig \ref{fig:broadcast}.  We can think of the information that is transmitted from the input transcription factor concentration $c$ to the output expression levels $\{g_i\}$ as a measure of the performance of the system, quantifying the power of the transcription factor to control the state of the cell.  This information $I(c;\{g_i\})$ is limited by the maximum absolute concentration of the input molecules and by the total number of messenger RNA (mRNA) molecules that can be transcribed.  We will assume that this total is large enough that the system as whole is not especially noisy---by looking at the expression levels $\{g_i\}$ we (or the cell) could infer the input signal $c$ with high precision.  The question is how the resources should be distributed. If there are a small number of target genes then each of them will be controlled precisely, but if the same total number of mRNA molecules is distributed across many targets then each of them will be very noisy.  We will show that, in a well-defined limit, the capacity for information transmission is optimized by using the maximum possible number of targets, so that each expression level is highly variable.

It is common to think that if elements are noisy then cells or organisms must average over many of these elements to achieve reliable functions.  The surprise here is that with a limited (but large) set of molecular resources, it actually is {\em better} to use many noisy elements than a few reliable ones.  It is not that we need to average in order to collect enough information, but rather that to maximize information the signal should be distributed among many noisy genes:  noisiness is a solution to the optimization problem.

\begin{figure}
\includegraphics[width=\linewidth]{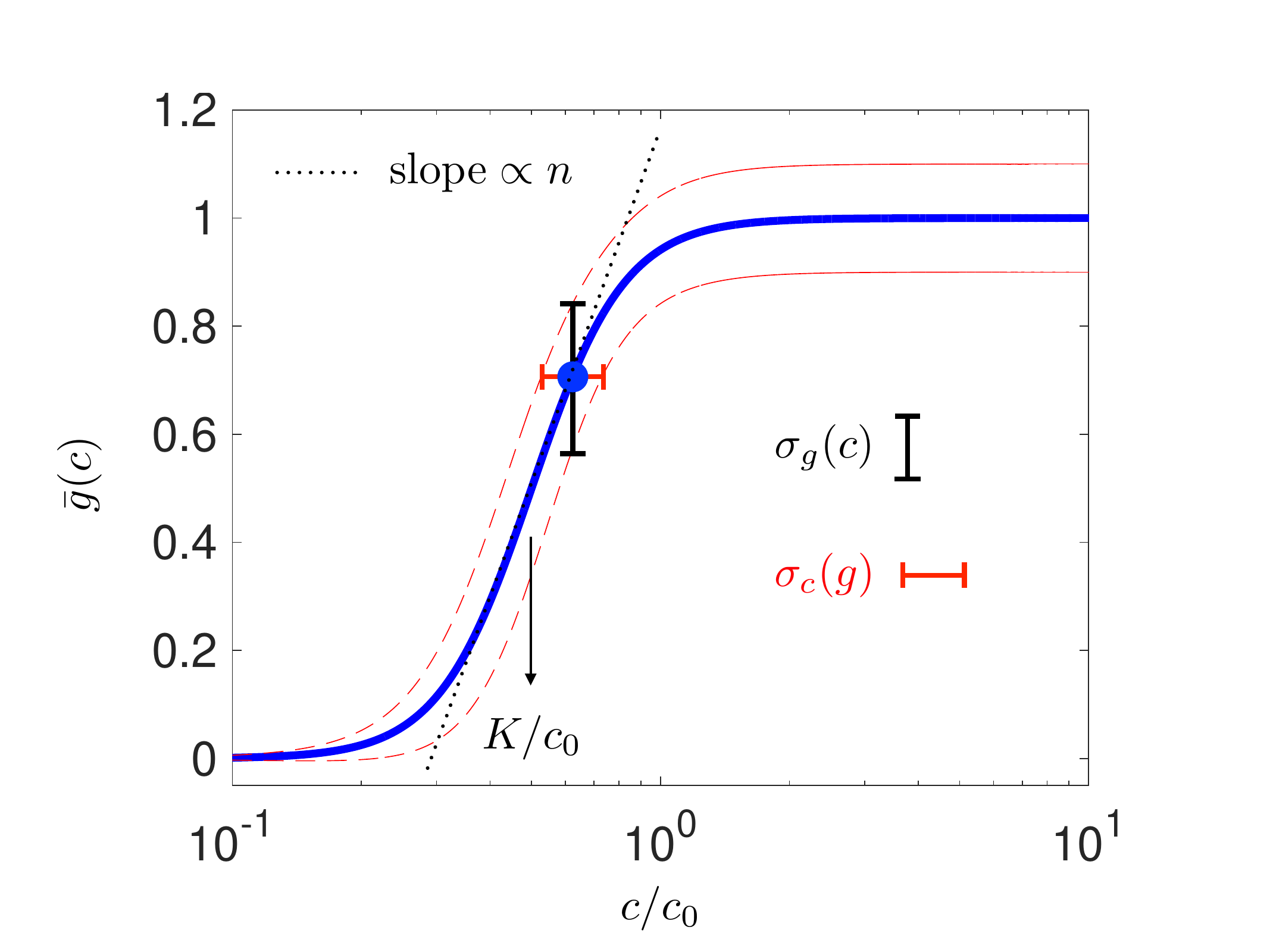}
\caption{Schematic input/output relation and noise associated with one arrow in Fig \ref{fig:broadcast}.  Mean expression level $\bar g (c)$ as a function of the transcription factor concentration $c$ (solid blue) $\pm$ one standard deviation $\sigma_g(c)$ (dashed red).  Expression is normalized so the maximal mean value is $\bar g = 1$ and TF concentration is in units of $c_0$ from Eq (\ref{c0}).  We mark the scale of half--maximal activation $K/c_0 = 0.5$ and the sensitivity $n=4$ from Eq (\ref{hill}). Orthogonal error bars show how noise in expression (black) is equivalent to an error in input concentration $\sigma_c(g)$ (red). \label{Hill+noise}}
\end{figure}

Once we have many target genes we also have many parameters.  Behind each arrow $c\rightarrow g_i$ is a schematic dependence of the expression level on the input transcription factor (TF) concentration (Fig \ref{Hill+noise}).  Such dependencies are defined (at least) by a characteristic concentration scale $K_i$ and a sensitivity $n_i$.  In a simple model we can think of these parameters as the affinity and cooperativity for binding of the TF to the relevant enhancer sites \cite{bintu+al_2005a,bintu+al_2005b}, although we need not take this literally.  We will see that at large $M$, where information transmission is optimized, the dependence of $I(c;\{g_i\})$ on the parameters $\{K_i,n_i\}$ is ``sloppy'' \cite{brown+sethna_03,gutenkunst+al_07,transtrum+al_15,quinn+al_22}, so that there are soft modes in parameter space along which combinations of parameters can vary with little impact on function.  This means that ensembles of cells could achieve nearly optimal information transmission with many different parameter settings, allowing the coexistence of optimality and variability, as  in other problems \cite{Bauer+al_2025}.

There are examples of TFs that operate in something like the broadcasting mode schematized in Fig \ref{fig:broadcast}, and in different contexts these have been labeled both ``master''  and ``promiscuous'' regulators \cite{Ohno1979SexDeterminingGenes, Pougach2014DuplicationPromiscuousTF},  but we suspect that the scenario we explore here is  too simple to be a realistic model for these particular systems. However, our approach provides proof of principle that noisiness and variability, rather than being evidence against optimization, can in fact emerge as consequences of an optimization principle.

\section{Formulating the problem}

Information transmission is limited by noise from the random behavior of individual molecules.  Given a maximum total number of molecules there is a bound on how much information can be represented or transmitted, but it is not obvious how to distribute these molecular resources.  Should cells use as many molecules as possible to make a single signal reliable, or distribute them across many noisy channels?  We want to address this problem in the context of transcriptional control, and we use what we believe is the simplest setting, the broadcasting network in Fig~\ref{fig:broadcast}.

Broadcasting is defined as a two-layer network in which inputs control a set of noninteracting outputs in feed-forward fashion. Here, we focus on the case where a single input drives $M$ outputs.  Concretely, we imagine a transcription factor that controls, independently, the expression of multiple genes. Limitations on molecular resources are set by the concentration $c$ of the TF having a maximum $c_{\rm max}$, and by fully expressed genes being encoded by $N_{\rm max}$ mRNA molecules.  We include only the minimal noise sources from random arrivals of the TF molecules at their binding sites \cite{Berg+Purcell_1977,Bialek+Setayeshgar_2005} and the Poisson fluctuations in mRNA synthesis and degradation \cite{paulsson2005stochastic}. In addition, we work in steady state, where concentrations are numbers rather than functions of time. Our analysis of information transmission in this system follows earlier work \cite{Tkacik+al_2009} but pushes into a new regime with unexpected results.  The discussion in this section is largely a review, included for completeness and to establish notation.

Before we begin the technical discussion we note that there are three layers of optimization even in this simple problem.  
First, if we think of the regulatory mechanisms as fixed, along with their noise levels, then we can adjust the distribution $P_{TF}(c)$ of input concentrations to maximize information transmission \cite{tkacik+al_2008a,tkacik+al_2008b}; the information that is transmitted at this optimum is referred to as the information capacity of the network \cite{Cover+Thomas_1991}.\footnote{There are two quantities that can be described as the information capacity.  The first, which we use here, is the maximum mutual information between input and output that can be achieved by adjusting the distribution of inputs.  The second, which is more relevant for engineered systems, is the maximum number of bits (or bits/sec) that can be transmitted without error.  The fact that these quantities are the same for many communication channels is one of the foundational results of information theory \cite{Shannon_1948,Cover+Thomas_1991}.} 
Second, we can adjust the parameters of the network to maximize information capacity with fixed limits on molecular resources \cite{Tkacik+al_2009}.
Finally, with $M$ target genes and $N_{\rm max}$ mRNA molecules for each, the cell has a total budget of $N_{\rm tot} = MN_{\rm max}$ molecules, and we can ask how best to distribute these resources: is information  maximized by a small number of reliable genes (small $M$, large $N_{\rm max}$) or a large number of noisy genes (large $M$, small $N_{\rm max}$)?  To solve the first two layers of optimization we follow previous work \cite{tkacik+al_2008a,tkacik+al_2008b,Tkacik+al_2009}.

We begin with definitions of the relevant probability distributions.  The input drives the output, so that if we know the TF concentration $c$ then there is some distribution of expression levels $P(\{g_i\}|c)$ that results.  The statement that the target genes do not interact is formalized by saying that this distribution factorizes,
\begin{equation}
P(\{g_i\}|c) = \prod_{i=1}^M P_i (g_i|c ).
\end{equation}
If the TF concentrations are drawn from a distribution $P_{TF}(c)$ then the probability (density) to find a combination of inputs and outputs is
\begin{equation}
P(\{g_i\}, c) = P(\{g_i\}|c)P_{TF}(c)
\end{equation}
and the distribution of output expression levels
\begin{equation}
P(\{g_i\}) = \int dc\, P(\{g_i\}, c) = \int dc\, P(\{g_i\}|c)P_{TF}(c).
\end{equation}
In addition to thinking of the input as driving the output, we can also think of the output as an encoding of the input, and there is a distribution $P(c|\{g_i\})$ of input TF concentrations that is consistent with a measured set of expression levels. Bayes' rule tells us that
\begin{equation}
P(c|\{g_i\}) = {{P(\{g_i\}, c)}\over{P(\{g_i\})}} = {{P(\{g_i\}|c)P_{TF}(c)}\over{P(\{g_i\})}} .
\end{equation}
We will work in the limit where the total amount of information is large, so that patterns of expression point sharply to a small range of TF concentrations around some most likely value $c_*(\{ g_i\})$.  Then we can approximate
\begin{equation} \label{conditional_P_c}
    P(c|\{g_i\})=\frac{1}{\sqrt{2\pi\sigma_c^2(\{g_i\})}}
    \exp{\bigg\{}-\frac{[c-c_*(\{g_i\})]^2}{2\sigma_c^2(\{g_i\})} {\bigg\}},
\end{equation}
where $\sigma_c(\{g_i\})$ is the error in estimating $c$ from $\{g_i\}$.

The mutual information between the input and outputs is defined, as usual \cite{Shannon_1948,Cover+Thomas_1991}, as
\begin{eqnarray}
    I(c;\{g_i\})&=&\int dc  \,d^Mg\, P(\{g_i\},c)\log\left[ \frac{P(\{g_i\},c)}{P_{TF}(c)P(\{g_i\})} \right] \nonumber \\
    &&\\
    &=& S[P_{TF}(c)] - \langle S[P(c|\{g_i\})]\rangle_{\{g_i\}} ,
\end{eqnarray}
where we identify  $S[P(x)]$ generally as the entropy of the distribution $P(x)$,
\begin{equation}
S[P(x)] = -\int dx\, P(x) \log  P(x) .
\end{equation}
We see from Eq (\ref{conditional_P_c}) that $P(c|\{g_i\})$ is approximately Gaussian, and we can compute the entropy of a Gaussian analytically, so that
\begin{equation}
S[P(c|\{g_i\})] = {1\over 2}\log [2\pi e \sigma_c^2(\{g_i\})] .
\end{equation}
Working consistently in the small noise limit, we can trade averages over the output for averages over the input:
\begin{eqnarray}
\langle f(\{g_i\})\rangle_{\{g_i\}} &\equiv& \int d^M g\, P(\{g_i\}) f(\{g_i\})\\
&\approx &   \int dc\, P_{TF}(c) f(\{g_i\}){\bigg |}_{\{g_i = \bar g_i(c)\}}. 
\end{eqnarray}
For compactness we will write
\begin{equation}
\sigma_c^2(\{g_i\}){\bigg |}_{\{g_i = \bar g_i(c)\}} = \sigma_c^2 (c).
\end{equation}
Putting the pieces together we have
\begin{eqnarray}
   I(c;\{g_i\}) &=& -\int dc\, P_{TF}(c) \log  P_{TF}(c) \nonumber\\
   &&\,\,\,\,\, - {1\over 2} \int dc\, P_{TF}(c) \log \left[ 2\pi e \sigma_c^2(c)\right] .
   \label{Igc_PTF}
\end{eqnarray}

The first layer of optimization is to choose the distribution of inputs.  We introduce a Lagrange multiplier $\alpha$ to fix the normalization of $P_{TF}(c)$, so we have to solve the variational problem
\begin{equation}
{\delta\over{\delta P_{TF}(c)}} \left[ I(c;\{g_i\}) - \alpha \int dc\, P_{TF}(c)\right] =0.
\end{equation}
Substituting from Eq (\ref{Igc_PTF}) and solving we find \cite{tkacik+al_2008a}
\begin{equation}
P_{TF}^*(c) = {1\over {\cal Z}} {1\over {\sigma_c(c)}} ,
\label{PoptTF}
\end{equation}
where the partition function
\begin{equation}
{\cal Z} = \int_0^{c_{\rm max}}  {{dc}\over {\sigma_c(c)}}.
\label{Zsigma_c}
\end{equation}
This yields the compact result for the optimal information transmission,
\begin{equation}
I_{\rm opt} = \log_2\left( {{\cal Z}\over{\sqrt{2\pi e}}} \right) \, {\rm bits}.
\label{IZ}
\end{equation}

The second layer of optimization is to choose the parameters of the network.  We see from Eqs (\ref{Zsigma_c}) and (\ref{IZ}) that these parameters can enter only through the effective input noise level $\sigma_c(c)$.
If the mean expression level of each gene is $\bar g_i(c)$ and the noise level in expression is defined by a standard deviation $\sigma_{g_i}(c)$, then for a single gene Fig \ref{Hill+noise} shows the usual propagation of errors
\begin{equation}
{1\over {\sigma_c(g)}} = {\bigg |} {{d\bar g (c)}\over{dc}}{\bigg |} {1\over {\sigma_g(c)}} ,
\end{equation}
although again this makes sense only if the noise is small.  If we can use the expression levels of multiple genes to infer the input TF concentration $c$, then these signals combine to reduce the effective noise level \cite{BevingtonRobinson2003DataReduction}
\begin{equation}
{1\over {\sigma_c^2(\{g_i\})}} = \sum_{i=1}^M {\bigg |} {{d\bar g_i (c)}\over{dc}}{\bigg |}^2 {1\over {\sigma_{g_i}^2(c)}} .
\label{errprop}
\end{equation}

To proceed, we need to specify the form of the input/output relations $\bar g_i (c)$ and the sources of noise in the network.  We make use of a phenomenological description, the Hill function \cite{Hill_1910}
\begin{equation}
    \Bar{g}_i(c)=\frac{c^{n_i}}{c^{n_i}+K_i^{n_i}}.
    \label{hill}
\end{equation}
We can imagine that $n_i$ molecules of the transcription factor bind cooperatively to sites close to the target, and that the expression level is proportional to the probability that all these sites are occupied; in this view $K_i$ is the affinity of the sites for the TF \cite{bintu+al_2005a,bintu+al_2005b}.  As far as we can tell this mechanistic view is not essential for what follows, and one can think of the Hill function as a prototypical monotonic function characterized by a scale $K$ along the concentration axis and a sensitivity or slope $n$ (Fig \ref{Hill+noise}). For simplicity, we consider only activators ($n>0$), but this is not essential to the results that follow.

We have chosen to measure expression in normalized units, so that the maximum mean value in Eq (\ref{hill}) is one.  If this maximum mean expression involves $N_{\rm max}$ mRNA molecules we expect that there will be Poisson fluctuations as these molecules are transcribed and degraded,  contributing a variance proportional to the mean expression,
\begin{equation}
\sigma_{\rm Poisson}^2 = {1\over {N_{\rm max}}} \bar g (c).
\end{equation}
In addition, there is noise as the TF molecules arrive at their binding sites, as first emphasized by Berg and Purcell \cite{Berg+Purcell_1977}.  We can think of this as a variance in the concentration that the regulatory mechanism ``sees,''
\begin{equation}
\langle (\delta c)^2\rangle = {c\over{ D a \tau}},
\label{BP}
\end{equation}
where $D$ is the diffusion constant, $a$ is the linear size of the binding sites, and $\tau$ is the effective integration time for expressed molecules.  This will make a contribution to the variance that we can calculate by propagating errors,\footnote{This is not quite right \cite{Kaizu+al_2014}, but should be a good enough approximation to help us understand the essential tradeoffs consdiered here.}
\begin{equation}
\sigma_{\rm BP}^2 =  {c\over{ D a \tau}} {\bigg |} {{d\bar g (c)}\over {dc}}{\bigg |}^2 .
\end{equation}
Putting these terms together we have
\begin{equation} 
\label{output variance}
    \sigma_{g_i}^2(c)=\frac{1}{N_{\text{max}}} \left[ \Bar{g_i}(c)+c_0c\left(\frac{d\Bar{g_i}(c)}{dc}\right)^2 \right]
\end{equation}
where 
 \begin{equation}
 c_0 ={{ N_{\rm max}}\over{Da\tau}}
 \label{c0}
 \end{equation} 
 is a natural concentration scale; the optimization problem is interesting because $c_0$ is comparable to the real concentrations of TFs \cite{Tkacik+al_2009}.

Since the information capacity is a monotonic function of the partition function $\cal Z$, we can take $\cal Z$ itself 
as the objective function to be optimized. From Eqs (\ref{Zsigma_c}), (\ref{errprop}), and (\ref{output variance}), we can write $\mathcal{Z}$ in dimensionless form as
\begin{equation}
    {\cal Z} =\sqrt{N_{\text{max}}}\bigintsss_0^C dx \left[\sum_{i=1}^M \frac{\left(d\Bar{g_i}(x)/dx\right)^2}{\Bar{g_i}(x)+x\left(d\Bar{g_i}(x)/dx\right)^2}\right]^{1/2}
    \label{dimless_Z} 
\end{equation}
where $x=c/c_0$, $C=c_{\text{max}}/c_0$, and the natural parameters are   $\{K_i/c_0,n_i\}$.  This is the optimization problem addressed in previous work \cite{Tkacik+al_2009}, and we follow their path in maximizing $\cal Z$ numerically.  What is new is that we push to larger numbers of genes $M$.

\section{Many Noisy Channels}

The partition function $\cal Z$ in Eq (\ref{dimless_Z}) depends on the parameters  $\{K_i/c_0,n_i\}$, and we can optimize over these numerically.\footnote{We use the MATLAB function \emph{fmincon}, and fix $n>0$ so that we are describing a TF that functions consistently as an activator of all its targets.  Over a wide range of $C$ we achieve good convergence of the optimization up to $M=30$, beyond which the algorithm slows substantially.}  
The resulting optimum ${\cal Z}_{\rm opt}$ still depends on the number of genes $M$, the maximum concentration of the inputs $C=c_{\rm max}/c_0$, and the maximum number of mRNAs for each gene $N_{\rm max}$.  The explicit dependence on $N_{\rm max}$ is simple, ${\cal Z} \propto\sqrt{N_{\rm max}}$, but there is also an implicit dependence through the concentration scale $c_0$ [Eq (\ref{c0})].

\begin{figure*} 
        \includegraphics[width=\linewidth]{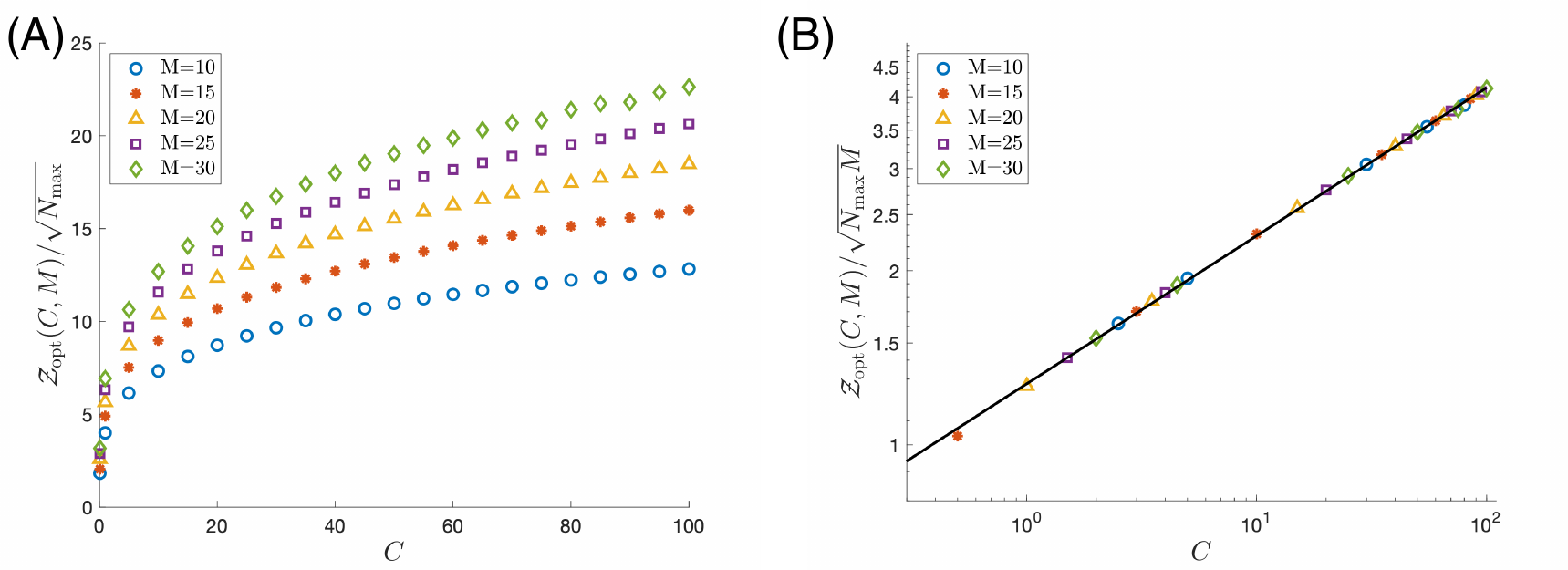}
    \caption{The partition function as a function of the dimensionless maximal input concentration. 
   (A) $\mathcal{Z}_{\rm opt}(C,M)/\sqrt{N_{\rm max}}$ for networks with different numbers of genes. (B) $\mathcal{Z}_{\rm opt}(C,M)/\sqrt{N_{\rm max}M}$ for each network, shown on a log-log scale. The curves collapse onto  $\mathcal{Z_{\rm opt}}\propto \sqrt{M} C^{1/4}$.
    \label{scaling}}
\end{figure*}

Results for ${\cal Z}_{\rm opt} (C,M)/\sqrt{N_{\rm max}}$ are shown in Figure \ref{scaling}A.  We see that the dependence on both $M$ and $C$ is monotonic, as expected:  with more molecules available the system can always transmit more information.  Results at different $M$ seem to have the same dependence on $C$, and this is made precise by rescaling in Fig \ref{scaling}B.

\begin{figure}[b]
    \centering
    \includegraphics[width=\columnwidth]{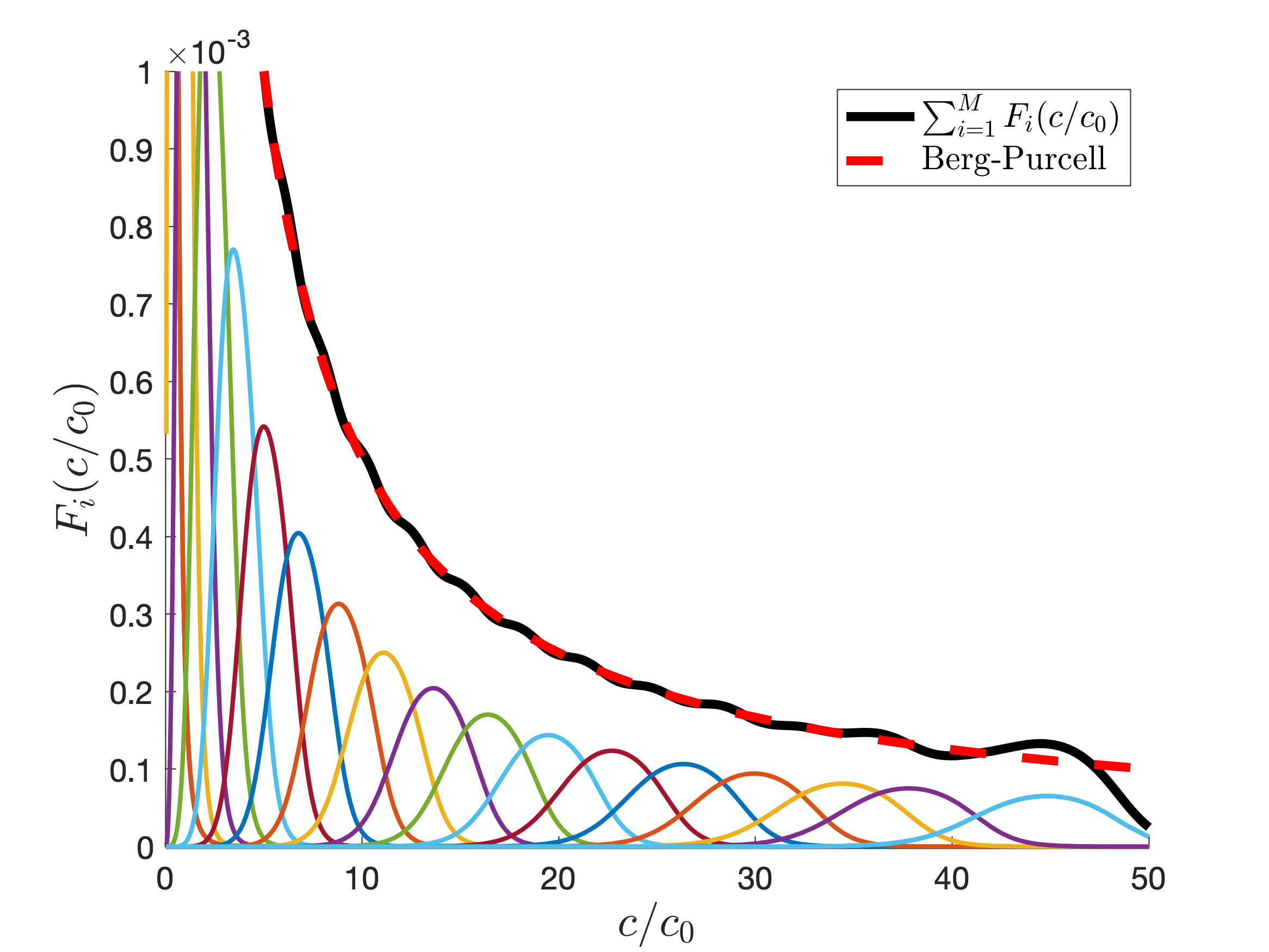}
    \caption{Tiling the concentration axis.  Functions $F_i(c/c_0)$ from Eq (\ref{Fix}); parameters $\{K_i,n_i\}$ are set to their optimal values for $M=20$ and $C=50$ (thin colored curves). The sum (solid black line) determines the effective noise level $\sigma_c(c)$ from Eq (\ref{sumF}). For comparison, if the effective noise level is given by the Berg-Purcell limit in Eq (\ref{BP}), we expect the sum to be proportional to $1/c$ (dashed red line).
\label{F+P}}
\end{figure}

The first result from Fig \ref{scaling} is that ${\cal Z}_{\rm opt} \propto \sqrt{M}$.  This is what we would expect if each of the terms inside the brackets of Eq (\ref{dimless_Z}) made the same contribution.  This suggests that the optimal choices of $\{K_i, n_i\}$ serve to distribute the contribution of these terms along the $c$ axis in some uniform way.
To see this more explicitly we can define
\begin{equation}
F_i (x) = \frac{\left(d\Bar{g_i}(x)/dx\right)^2}{\Bar{g_i}(x)+x\left(d\Bar{g_i}(x)/dx\right)^2},
\label{Fix}
\end{equation}
so that
\begin{equation}
    {\cal Z} =\sqrt{N_{\text{max}}}\bigintsss_0^C dx \left[\sum_{i=1}^M F_i(x) \right]^{1/2} .
\end{equation}
In Fig \ref{F+P} we show the functions $F_i(x)$, with all the parameters $\{K_i,n_i\}$ set to their optimal values, for $M = 20$ and $C=50$.  

We see that the different $F_i(c/c_0)$ tile the concentration axis, increasing their spacing and increasing their width as their centers moves toward larger concentrations.  This spacing makes it plausible that each term can make a similar contribution to the integral that defines $\cal Z$.  If we add the terms
\begin{equation}
{1\over{N_{\rm max}}}{1\over{ \sigma_c^2(c)}} = \sum_{i=1}^M F_i(x) ,
\label{sumF}
\end{equation}
we find that $\sigma_c^2\propto c$ over most of the dynamic range, as expected if the use of multiple target genes allows the effective noise level to become proportional to the Berg--Purcell limit from Eq (\ref{BP}).

At large $M$ it is useful to think about a continuum limit, with the sum over genes becoming an integral over the binding constants $K_i$,
\begin{equation}
\sum_{i=1}^M F_i(x; K_i/c_0 , n_i ) \rightarrow M \int_0^C dk\,\rho(k) F(x; k, n_*(k)),
\end{equation}
where we make explicit that $F(x)$ also depends on $K/c_0$ and $n$, $n_*(k)$ is the optimal $n$ for each $k$, and the density is normalized
\begin{equation}
\int_0^C dk\,\rho(k) = 1;
\end{equation}
we assume that once the network is optimized there are no binding constants outside the natural range $0\leq K \leq c_{\rm max}$.  The function $F$ peaks at $k\sim x$, so that each gene makes its dominant contribution near the point of half--maximal activation.  The peak height is
\begin{equation}
F_{\rm peak} (x;  k=x, n) \sim {{n^2}\over{x^2 + n^2 x}},
\end{equation}
and the peak width is $\Delta k \sim x/n$.  We recover simple scaling if $n_*(k)\sim \sqrt{k}$, which we observe from the numerics. This means that the width of $F_i(x)$ should be $\Delta x \sim  x_{\rm peak}^{1/2}$ and this is consistent with what we see in Fig \ref{F+P}.  Then if the density is smooth we can approximate
\begin{equation}
\int_0^C dk\,\rho(k) F(x; k, n_*(k)) \sim \rho(x) F_{\rm peak} (x)\Delta k \sim {{\rho(x)}\over \sqrt{x}} .
\end{equation}
To recover $\sigma_c^2\propto c$ from Eq (\ref{sumF}) then requires $\rho(x) \sim 1/\sqrt{x}$, and this again is consistent with the spacing of the optimal $F_i(x)$ in Fig \ref{F+P}.  But to keep $\rho(x)$ normalized we must have $\rho(x) = 2/\sqrt{Cx}$.  Finally, we put these terms together to give
\begin{equation}
{\cal Z}_{\rm opt} \sim \sqrt{N_{\rm max} M}\int_0^C dx\,\left[{{\rho(x)}\over \sqrt{x}}\right]^{1/2} \sim \sqrt{N_{\rm max} M}C^{1/4},
\label{C14}
\end{equation}
and this is what we see in Fig \ref{scaling}B.

Another way of understanding the result in Eq  (\ref{C14}) is that if we had just one concentration sensor that reached the Berg--Purcell limit, we would have ${\cal Z}_{\rm opt} \sim C^{1/2}$.  We have $M$ sensors, but these  have limited dynamic range and so must be distributed along the concentration axis to match the noise levels.  Since there are roughly $\sqrt{C}$ distinguishable concentration levels, the number of sensors ``looking'' at each concentration is $N_{\rm sens} \sim M/\sqrt{C}$.  Then we have
\begin{equation}
{\cal Z}_{\rm opt} \rightarrow \sqrt{N_{\rm sens}} C^{1/2} \sim \sqrt{M} C^{1/4},
\end{equation}
as in Eq (\ref{C14}).

For the third and final layer of optimization, we assume that the organism incurs a metabolic penalty to produce an mRNA molecule, such that a reasonable constraint is to allow for a total mRNA ``budget.'' If we hold this budget $N_{\text{tot}}=MN_{\text{max}}$ fixed, we can rewrite $\mathcal{Z}_{\rm opt}$ as
\begin{equation}
{\cal Z}_{\rm opt} \propto(N_{\text{tot}} M c_{\text{max}}Da\tau)^{1/4}
\end{equation}
Recalling that the maximal input concentration $c_{\text{max}}$ and biophysical parameters $D$, $a$, and $\tau$ are externally fixed, we see that, no matter what the total mRNA budget is, the best thing to do is always to increase the number of genes.   In principle, optimizing would push the network toward arbitrarily large $M$, but at some point our small noise approximations will break down. Nonetheless, the prediction  is that the optimal strategy is to increase the number of information transmitting channels, rather than to decrease the noise in any given channel.  Counterintuitively, perhaps, information transmission is optimized in this system when individual expression levels are noisy.

A corollary of these results is that the distribution of input concentrations and the distribution of binding constants are both strongly biased toward low values. It seems fair to say that, in the absence of a concrete calculation, one might not have expected the combination of low input concentrations and noisy outputs to be a consequence of optimizing information flow.

\section{Parameter variations}

Once we have a system where a single TF broadcasts to many genes, we also have a large number of parameters $\{K_i, n_i\}$.  One might worry that, while having many genes optimizes information transmission, achieving this optimum could require fine tuning of these many parameters.  To explore this we consider how information transmission behaves in the neighborhood of the optimum.

Let us label the parameters $\bm{\theta} \equiv \{K_i, n_i\}$.  Then there is an optimum setting $\bm{\theta} = \bm{\theta}^*$ that maximizes $I(c;\{g_i\})$, and near this setting we have
\begin{equation}
I(\bm{\theta}) = I_{\rm max} - {1\over 2}\sum_{n,m =1}^{2M} (\theta_n - \theta_n^*) \hat{H}_{nm} (\theta_m - \theta_m^*) + \cdots ,
\label{HD}
\end{equation}
where $\hat H$ is the Hessian matrix that describes the ``stiffness'' holding parameters near their optimal values.  In general the different parameters interact, so it is natural to look at the eigenvalues $\{\lambda_\mu\}$ and eigenvectors of the Hessian---eigenvectors define directions in parameters space that have independent effects on performance (to leading order), and eigenvalues measure this sensitivity.

\begin{figure}[t]
\includegraphics[width=\linewidth]{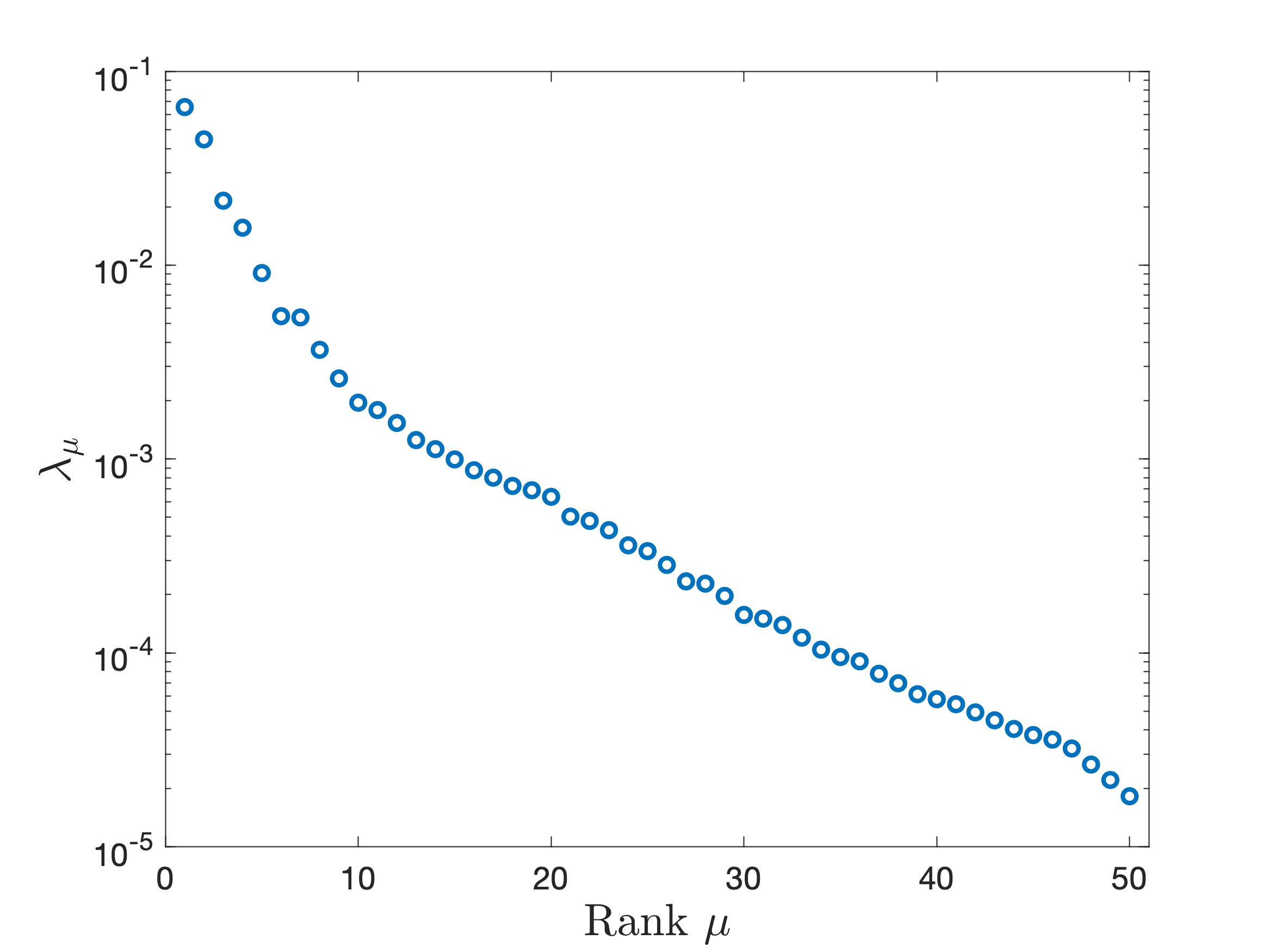}
\caption{Spectrum of the Hessian matrix. Eigenvalues $\{\lambda_\mu\}$ of $\hat H$ from Eq (\ref{HD}) in a model with $M=25$ genes and $c_{\rm max}/c_0 = 50$.  
\label{fig:eigenvalues}}
\end{figure}

Figure \ref{fig:eigenvalues} shows the spectrum of the Hessian with $M=25$ genes and $c_{\rm max}/c_0 = 50$; no other parameters enter.  Since $K$ is measured in units of $c_0$ it is dimensionless, as is $n$, all elements of $\hat{H}$ have the same units (bits).  We see that the eigenvalues span nearly four decades, with much of the spectrum forming an exponential tail.

An exponential tail  is equivalent to the statement that these eigenvalues are distributed uniformly on a logarithmic scale or more ambitiously that the spectrum is scale invariant.  This pattern is seen in many  multi--parameter fits to data on complex systems: some combinations of parameters are determined precisely, and others are not \cite{brown+sethna_03,gutenkunst+al_07}.  Sethna and colleagues have argued that this behavior defines a broad class of ``sloppy models'' \cite{transtrum+al_15}  and that this gives a path to model simplification \cite{quinn+al_22}.   

In our case, the parameters are physical properties of the genome, and there is no natural path to model simplification.  In particular, we can imagine evolution modifying individual values $K_i$ by changing the sequence to which the TF binds in the enhancer for gene $i$, and modifying the individual $n_i$ by duplicating or deleting binding sites, but there is no obvious way to remove some fractional superposition of these elements and ``simplify'' the network.

The sloppy spectrum does mean that cells can tolerate substantial variations in parameters, as emphasized recently in other contexts \cite{Bauer+al_2025}.   This invites us to think about an ensemble of cells with parameters chosen at random from a distribution $P(\bm{\theta})$.  We can ask for a distribution that allows parameters to be as random as possible while, on average, cells  convey some near--optimal  information $\langle I(\bm{\theta})\rangle$.  The distribution with these two properties is the Boltzmann--like distribution,
\begin{equation}
P(\bm{\theta}) = {1\over {Z(T)}} \exp\left[  I(\bm{\theta})/T\right],
\end{equation}
where the information plays the role of (negative) energy, varying the ``temperature'' $T$ trades between the degree of variation and the magnitude of information, and $Z(T)$ serves to normalize the distribution.  This picture of a Boltzmann--like functional ensemble of cells or organisms---exploring parameter space as widely as possible while maintaining some high average performance---has arisen in multiple contexts, with different interpretations \cite{Taylor+al_2007,Bodova+al_2016,Bodova+al_2021,DeMartino+al_2017,DeMartino+al_2018,Mlynarski+al_2021}.  In the approximation of Eq (\ref{HD}) we can compute both the average information transmission and the entropy $S$ of the distribution in parameter space
\begin{equation}
S = -\int d^{2M}\theta\, P(\bm{\theta}) \ln P(\bm{\theta}) ;
\end{equation}
each is a function of $T$ but they can be related directly to one another to give
\begin{equation}
\langle I (\bm{\theta}) \rangle = I (\bm{\theta}^*) - {{M\Lambda}\over{\pi e}}e^{S/M},
\end{equation}
where $\Lambda$ is the geometric mean of the Hessian eigenvalues $\{\lambda_\mu\}$ \cite{Bauer+al_2025}.

For the example in Fig \ref{fig:eigenvalues} we have $\Lambda = 3.86\times 10^{-4}$.  This means that we can have, on average, more than one bit of variation in each parameter and still achieve information transmission within $\sim 0.01\,{\rm bits}$ of the maximum. Thus optimization of information transmission with bounded molecular resources not only drives the system into a regime where each of the many target genes has noisy expression, but the parameters of each TF/DNA interaction can fluctuate substantially with hardly any loss of information.  We have optimality without fine tuning.

\section{Discussion}

Life depends on information as much as on energy.  In order to survive and reproduce, organisms need to know what to do and when to do it.  Just a decade after the introduction of information theory, Barlow proposed that part of the function of neural computation was to generate efficient representations of sensory information \cite{Barlow_1959,Barlow_1961}, and this  hypothesis remains a driver of thinking about neural coding more six decades later \cite{Manookin+Rieke_2023}.  Efficient representation means conveying as much relevant information as possible with limited resources, and perhaps this is even clearer in biochemical and genetic networks where the resources are molecules \cite{tkacik+al_2008a,tkacik+al_2008b}.  The early fruit fly embryo provides evidence for such efficient representation \cite{dubuis+al_2013b,petkova+al_2019,mcgough+al_2024}, and the theory of optimal information flow in these networks is now a substantial field \cite{Tkacik+tenWolde_2025},  From a physics point of view it is especially attractive that the same principles are being used at such different levels of biological organization \cite{Bialek_2012,Bialek_2024}.

For both neural and genetic networks, direct observations of noisiness and variability seem like {\em prima facie} evidence against ideas of optimization.  The result of our work is that this conflict can be resolved.  Maximizing the information transmitted by transcriptional regulation leads to a distribution of molecular resources across many genes, so that expression of each gene becomes very noisy.  The optimal distribution of transcription factor concentrations also is biased toward very low values.  Finally, the ``broadcasting'' network that maximizes information flow is enormously tolerant to parameter variations, so that we expect an ensemble of near--optimal cells to explore these parameters widely in the absence of other constraints.

\begin{acknowledgments}
We thank M~Nikoli\'c for helpful discussions. Early stages of this work were supported in part by the National Science Foundation through the Center for the Physics of Biological Function (PHY--1734030).
\end{acknowledgments}

\bibliography{NL+WB_25.bib}

\end{document}